\documentclass[12pt]{article}
\usepackage[T1]{fontenc}
\usepackage{graphicx}
\usepackage{bookman}
\usepackage{latexsym}
\usepackage{amssymb}
\usepackage{amsbsy}
\usepackage{cite}
\begin{document} 
  \title{Adiabatic decaying vacuum model for the universe}
 \author{ M.de Campos$^{(1)}$}
\maketitle
\footnote{
{ \small \it $^{(1)}$ UFRR\vspace{-0.3cm}}\\

{\small \it Campus do Paricar\~ana.  Boa Vista. RR. Brasil. \vspace{-0.3cm} }\\
                  
{\small campos@dfis.ufrr.br }} 
   
   \begin{abstract}
 We study a model that the entropy per particle in the universe is constant.  The sources for the entropy
are the particle creation and a $\Lambda$ decaying term.  We find exact solutions for the Einstein field equations and show the compatibility
of the model with respect to the age and the acceleration of the universe.
\\
\\
Key words - age of the universe, accelerated universe, entropy of the universe.
\end{abstract}
\section{Introduction}
Only few years after introducing the field equations of General Relativity, Albert Einstein includes in the field equations the cosmological constant.  Moreover,
in the light of the experimental evidence of the expansion of the universe, himself considered the inclusion
of the cosmological constant ``...the biggest blunder of my life.''\cite{Gamov}.

Recently, the experiments that use supernova as standard candles, realized independently by two groups
\cite{Perlmuter}, \cite{Riess} indicates that our universe is accelerated. The increase of the expansion velocity 
of the universe can be explained by a repulsive force that appears in the energy momentum tensor as 
a negative pressure. This negative pressure can be supply by the cosmological constant or term, time dependent.

Before the supernova indications that our universe is accelerated, L. Krauss and M. Turner \cite{Krauss} call our
attention that
 `` The Cosmological Constant is Back ''.  They cited the age of the universe, the formation of
 large-scale structure and the matter content of the universe as the data that cry out by a cosmological constant.

In the another hand, the inclusion of the cosmological constant creates new problems.  Some of them are old
, as the problem about the discrepancy among the observed value for the energy density for the vacuum,
and the large value suggested by the particle physics models \cite{Weinberg}, \cite{Garrig}.  This small value
for the cosmological constant today can be justify if we admit a cosmological term ($\Lambda $) 
, dependent of time,  in the place of the cosmological constant.  The $\Lambda $ term represents a type of 
dark non barionic matter that differs from the usual because it is not gravitationally clustered at all
scales.  The inclusion of the $\Lambda $ term  results in a cosmological
scenario with a good agreement in respect to the estimative for the age of the universe, the anisotropy of the 
cosmic microwave background radiation, and the supernova experiments \cite{Carroll}, \cite{Sahni}.

In another hand, models that includes particle production in the cosmological scenario also produce
compatible results in respect to the experimental evidence resulting in a relative attention to this model
in the literature \cite{Campos}, \cite{Lima1}, \cite{Abramo}, \cite{Calvao}, \cite{Harko}, \cite{Lima}, \cite{Prigogine} .

In this work, we study models with a dissipative pressure due to decaying of a cosmological term
and  particle production. The dissipative pressure is related to the decaying of $\Lambda $
 and to particle production using the adiabatic criterious, $\dot{\sigma} = 0$, where $\sigma =\frac{S}{n} $.
$S$ is the entropy and $n$ is the particle density. The adiabatic condition is necessary for to maintain the 
compatibility with the constraints of the cosmic microwave background radiation when the creation of photons
are considered \cite{Lima}.  
\section{The model}
\subsection{Adiabatic criterious}
Taking into account the time derivative of the energy conservation equation
 \begin{equation}
nTd\sigma = d\rho -(\mu + T\sigma)dn
\end{equation}
and the Euler equation that defines the chemical potential $\mu$
\begin{equation}
\mu = \frac{\rho + P}{n} - T\sigma \, ,
\end{equation}
we obtain
\begin{equation}
nT\dot{\sigma} = \dot{\rho }-\frac{\rho + P}{n}\dot{n}\, .
\end{equation}

Constructing the entropy four-vector $S^{\alpha} = n\sigma u^{\alpha}$, 
we can express the second law of thermodynamics in a covariant way
\begin{equation}
S^{\alpha}_{;\alpha} = n\dot{\sigma} + \sigma \Psi \, , 
\end{equation}
where we use the continuity equation 
 \begin{equation}
\dot {n} + n \Theta = \Psi\, .
\end{equation}
The $\Psi $ denotes the source of particles and  $\Theta$ is the expansion ($\Theta = 3\frac{\dot{R}}{R}$, where $R$
is the scale factor).

The energy momentum-tensor is described by a perfect fluid with a term that can be consider as the energy
of the vacuum, namely
 \begin{equation} 
T^{\alpha \beta} = (\rho + P_T)u^{\alpha }u^{\beta } - P_T g^{\alpha \beta} +\Lambda g^{\alpha \beta}\, .
\end{equation}
The $P_T$ is the total pressure that includes the usual thermodynamic pressure ($P$)
 and a dissipative pressure ($\Pi$ ), $\rho $ is the matter-energy density and $u$ is the quadri-velocity vector.
For this energy-momentum tensor the expression for the energy conservation, $u^{\alpha }T^{\alpha \beta}_{;\beta} =0$, 
results 
\begin{equation}
\dot{\rho}+(\rho +P_T)\Theta = -\frac{\dot{\Lambda}}{8\pi G}\, .
\end{equation}
Using expressions (3), (5) and (7) we obtain for the entropy relation, equation (4), the expression
\begin{equation}
S^{\alpha}_{;\alpha} = \frac{n}{T}\{-\frac{\dot{\Lambda}}{8\pi G}-(\rho +P)\frac{\Psi}{n}-\Pi\Theta\}+\sigma\Psi \, .
\end{equation}
Taking into account the decay of $\Lambda$ and the particle production as an  adiabatic process ($\dot{\sigma} = 0$), consequently
\begin{equation}
\frac{\dot{\Lambda}}{8\pi G}+\frac{(\rho +P)}{n}\Psi + \Pi \Theta = 0 \, .
\end{equation}
\subsection{Model with a cosmological term and particle production}
We consider the space-time described by the metric
\begin{equation}
ds^2 = dt^2-R^2(dr^2+r^2d\theta ^2+r^2(\sin(\theta))^2d\phi^2)\, .
\end{equation}
The field equations for the space time described above and the energy momentum tensor (6)
are
\begin{eqnarray}
2\frac{\ddot R}{R}+\frac{{\dot R}^2}{R^2}&=& -8\pi G(P+\Pi)+\Lambda \\
3\frac{{\dot R}^2}{R^2}&=& 8\pi G \rho +\Lambda
\end{eqnarray}
Eliminating $\rho $ from the field equations (11), (12) and using the expression (9) for the 
dissipative pressure, we obtain
\begin{equation}
2\frac{\ddot R}{R}+({\frac{\dot R}{R}})^2[1+3\nu-\frac{3(\nu +1)\Psi}{n\Theta}] =
\frac{\dot \Lambda}{\Theta} + \Lambda(1+\nu)(1-\frac{\Psi}{n\Theta})\, ,
\end{equation}
where $\nu $ is a constant given by the state equation $P = \nu \rho$.

The particle production is due to decaying of $\Lambda$. So, we consider that the contribution of the $\Lambda$ term
and particle production to the dissipative pressure, are proportional.  Namely,
\begin{equation}
\frac{\dot \Lambda }{8\pi G} = w(\rho +P) \frac{\Psi }{n}\, ,
\end{equation}
$w $ is a constant.  Note that, $\Psi =0 $
implies in the entropy of the universe and $\Lambda$ constants.

Using equation (9) (14) the dissipative pressure is given by
\begin{equation}
\Pi = -(1+w)\frac{\Psi }{n \theta }(\rho + P)\, .
\end{equation}
Carvalho and Lima \cite{Carvalho} call the attention that the "ansatz" used by
Chu and Wu \cite{Wu} do not fix $\Lambda \propto R^{-2}$.  Therefore, it is 
possible a different decaying law for $\Lambda$.
Assuming $\Lambda = \frac{1}{l_{pl}^2} [\frac{t_{pl}}{t_H}]^n $, where $l_{pl}$, $t_{pl}$ and $t_H$ are respectively the Planck
length and time, the Hubble time where  n is a integer.  To get rid the constant of Planck dependence of $\Lambda$ we must have 
$n=2 $ \cite{Carvalho},\cite{Wu}, \cite{Pad}.
The Hubble time is proportional to $H^{-1}$,
then we can consider the scaling of $\Lambda $ as
\begin{equation}
\Lambda = \alpha H^2 \, .
\end{equation}
Taking into account relations (7), (9) and (12) we find
\begin{equation}
(\rho + P)\theta (1-\frac{\Psi }{n\theta })= \frac{1}{8\pi G}{(\Lambda - 3H^2)}^{\cdot } \, .
\end{equation}
Substituting (16) into (17), we find the particle source corresponding to the $\Lambda $
term, namely
\begin{equation}
\Psi = \beta n \theta \, ,
\end{equation}
where $\beta $ is a constant.  This source of particle production has appear in the literature, 
where the constant $\beta$ is described by a phenomenological constant    \cite{Campos}, \cite{Lima1}, \cite{Lima}, \cite{Bel}.
In our case this constant is given by   $\beta = \frac{1}{1+w(1-\frac{3}{\alpha})}$.

The differential equation that rules the model can be obtained substituting (17) and (18)
into the equations (11) and (12), resulting in the field equation
\begin{equation}
2\frac{\ddot R}{R}+K{(\frac{\dot R}{R})}^2=0 \, ,
\end{equation}
where $K=1+3\nu+\frac{3\alpha (1+\nu)}{3w-\alpha (1+w)}$.
Integration of this equation results
\begin{equation}
R(t) \propto t^{\frac{2}{2+K}} \, .
\end{equation}
This solution is not valid for $K = -2$ that implies $ w = 0 $, but
if we put $w =0 $ into equation (13) we obtain the model with $\Lambda $ constant.
Considering $\Lambda = 0 $ in equation (13) we recover the open system cosmological model \cite{Lima}, \cite{Prigogine}.
\section{Observational Constraints}
In this work  we consider the entropy per particle constant ( $\dot \sigma = 0 $).  Using expressions (4), (8) and taking into account 
that the $\Lambda$ term decays, we obtain
\begin{equation}
S^{\alpha}_{;\alpha} > 0  \longrightarrow \Psi >0 \, .
\end{equation}
To consider $\dot{\Lambda}<0 $ is plausible since we expected that $\Lambda$ have a greater value in the past.
However, for do not violate the second law of thermodynamics we must have particle creation and not
particle destruction.  
\subsection{The acceleration of the universe}
According recent observations from the supernova type IA, the expansion velocity of the universe increases.
Generally an negative pressure is responsible for the acceleration of the universe. In this work the dissipative pressure $\Pi$ is
the responsible for this.  Looking the expressions (9) and (14) we note that
\begin{equation}
-1 < w < 0 \, ,
\end{equation}
since that, the $\Lambda $ term decays.

In another hand, the  deacceleration parameter ($q = -\frac{R \ddot R}{\dot R ^2}$) is given by
\begin{equation}
q=\frac{K}{2}\, ,
\end{equation}
where $K = 1+\frac{3 \alpha}{3w - \alpha (1+w)}$.  So, to obtain an accelerated universe we must satisfy
\begin{equation}
1+\frac{3 \alpha}{3 w -\alpha (1+w)} < 0 \, .
\end{equation}
The profile of the deacceleration parameter (23) is given by:
\newpage
\begin{figure}[!ht]
\centerline{\includegraphics[width=6cm]{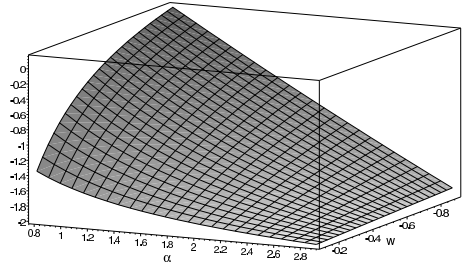}}
\caption{Evolution of the deacceleration parameter versus the constants $\alpha$ and $w$.}
\label{fig:figure 1}
\end{figure}
\subsection{The age of the universe}
The High Z supernova team estimates for the age of the universe $14.2 \pm 1.7$ Gyrs \cite{Riess} and $14.9 \pm 1.4 $ Gyrs \cite{Perlmuter}
.  Carreta et al. \cite{Carreta} using results from Hipparcos, RRlyrae and Chepheids to re-calibrate the globular clusters distance scale, found
 $12.9 \pm 2.9 $ Gyrs.  Butcher \cite{Butcher} used the abundance of the $Th^{32}  $ as a method for estimate the age of the stars and consequently a lower
limit for the age of the universe.  Using the relation among the  abundance of $Th $ and $Eu $, Westin et al. \cite{Westin} find about $15.0$ Gyrs
, while Johnson and Bolte \cite{Bolte} found about $11.4$ Gyrs.  Recently, Krauss and Chaboyer \cite{Chaboyer} estimates a level
lower limit on the age of the universe, $11$ Gyrs with $95 \% $ confidence.

In view of the above results we must to consider the possibility that we can not live in a flat matter dominated universe, since 
the universe in this scenario is younger than the experimental evidence indicates.

Using the scale factor (20), we find for the age of the universe the expression
\begin{equation}
t_0 = \frac{2}{3H_0}[1+\frac{2}{w(\alpha -3)}] \, .
\end{equation}
To obtain a universe older than the standard model is necessary that
\begin{equation}
1+ \frac{2}{w(\alpha -3)} > 0 \, .
\end{equation}
\newpage
The evolution for condition (26) have the following profile.
\begin{figure}[!ht]
\centerline{\includegraphics[width=6cm]{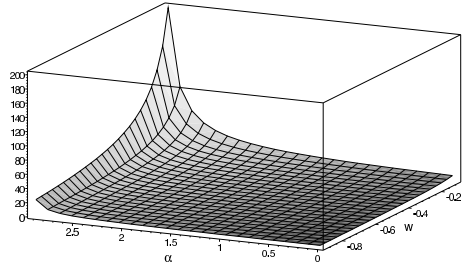}}
\caption{Evolution of condition (26) versus the constants $\alpha$ e $w$}
\label{fig:figure 2}
\end{figure}

Considering the interval of validity for $w$, the conditions (26) and (24), and the fact that we have particle creation ($\beta >0$), we can infer 
a validity range for $\alpha$, namely $1<\alpha<3$.
\section{Conclusions and final remarks}
We consider a model with particle production at the expenses of decay of a cosmological term and obtain an exact solution for the field equations.
Considering, the second law of thermodynamics, the dissipative pressure (15) and in the results from supernova type IA
, which indicates that we live in an accelerating universe, we obtain 
an interval for the constant $w$ that gives the contribution of $\Lambda $ term to the dissipative pressure. Namely
$-1<w<0$.  The indications that we live in a universe older then the stablished by the standard model
results in a validity range for $\alpha$ ($1<\alpha<3$).
Looking to the profiles, figure (1) and (2), in the validity intervals for $w$ and 
$\alpha $ the universe is accelerated and older than the estimatives from the standard model.

The next step for a future work is analyze the lensing probability for this model, 
luminosity distance diagram and the influence of this background in the diameter distance of distant objects.
Generally, the accelerated models for the universe presents the growing modes for the density contrast evolving more slower than the modes obtained
for the standard model. This topic is also motive for future study.

\end{document}